\documentclass[namedreferences]{SolarPhysics}
\usepackage[optionalrh]{spr-sola-addons} 
\usepackage{graphicx}                    
\usepackage{color}                       
\usepackage{url}                         

\usepackage[T1]{fontenc}

\newcommand{\aap}{{\it Astron. Astrophys.}}
\newcommand{\etal}{{\it et al.}}
\newcommand{\solphys}{{\it Solar Phys.}}
\newcommand{\apj}{{\it Astophys. J.}}
\newcommand{\an}{{\it Astron. Nachr.}}
\newcommand{\basi}{{\it Astron. Soc. Of India Bull.}}

\begin{document}

\begin{article}

\begin{opening}

\title{Determination of 3D Trajectories of Knots in Solar Prominences Using MSDP Data}

%
\author{Maciej~\surname{Zapi{\'o}r}\sep
        Pawe{\l}~\surname{Rudawy}
       }

%
\runningauthor{M.~Zapi{\'o}r, P.~Rudawy}
\runningtitle{Determination of 3D Trajectories of Prominence Knots}

%
\institute{M.~Zapi{\'o}r\sep P.~Rudawy \\ Astronomical Institute of the University of Wroc{\l}aw \\ 51-622 Wroc{\l}aw, ul. Kopernika 11, Poland\\
                     email: \url{zapior@astro.uni.wroc.pl}\newline
\newline
                     P.~Rudawy\\
                     email: \url{rudawy@astro.uni.wroc.pl}
             }

\begin{abstract}
In this paper we present a new method of restoration of the true thee-dimensional trajectories of the prominence knots based on ground-based observations taken with a single telescope, which is equipped with a Multi-Channel Subtractive Double Pass imaging spectrograph. Our method allows to evaluate true three-dimensional trajectories of the prominence knots without any assumptions concerning the shape of the trajectories or dynamics of the motion. The reconstructed trajectories of several knots observed in three prominences are presented.
\end{abstract}

\keywords{Prominences, Dynamics; Prominences, Formation and Evolution; Instrumentation and Data Management}

\end{opening}

\section{Introduction}

All solar prominences show motions of various spatial and temporal scales, ranging from small-scale and slow displacements of individual structures (knots) in quiet prominences up to violent eruptions of the whole structure in eruptive prominences. It is without doubt that investigation of the true three-dimensional (3D) trajectories of the individual knots of the evolving prominences can provide very valuable data concerning the spatial structure and dynamics of the prominences themselves and their magnetic skeletons, overall structure of huge magnetic systems encompassing prominences, as well as Coronal Mass Ejection-prominence relations. The observations and investigations of the magnetic fields in solar prominences are very difficult due to insufficient spatial resolution and low sensitivity of direct or indirect measurement methods of the magnetic fields currently used. For this reason the observations of the bulk flows along the magnetic ropes provide us with very valuable data.

Up to now most data concerning trajectories and kinematics of prominence material is based on analysis of filtergram observations (\textit{e.g.} images taken in various spectral lines, mostly in H$\alpha$ line) and spectral observations taken with classical spectrographs. Analysis of time-series of images provide parameters of the motion of the observed structure (its translation, velocity and acceleration) only in a plane of the sky (\textit{e.g.} \opencite{Rot1954}; \opencite{McCabe1970}; \opencite{Rompolt75}; \opencite{Kim1990}; \opencite{Auras1991}; \opencite{Vrsnak1993}; \opencite{Udin1994}; \opencite{Tandberg1995}; \opencite{Rudawy1998}) while 3D trajectories of the plasma (namely separated knots of material) could be evaluated only under certain strong and arbitrary assumptions. To do this Billing and Pecker assumed that the acceleration of the prominence knots is always normal relative to their velocity vector \cite{Billing54}, Palu\v s assumed spiral motions of the knots on conical surfaces \cite{Palus72}, Makhmudov and coauthors assumed constant acceleration of the falling knots \cite{makh80}, and Ballester and Kleczek introduced their own method based on an extended geometrical analysis of the data \cite{ball83}. In case of data taken with classical spectrographs, which have a very limited field of view and/or time resolution, it is possible to evaluate line-of-sight (LOS) velocities as well as transverse translations of the observed separated structure (using slit-jaw images), but relative slowness of the scanning process of the entrance slit across the observed prominence substantially reduces the temporal resolution of the observations. Due to the described limitations only a small volume of data collected with ground-based instruments related to the 3D trajectories and kinematics of the prominence plasma is available presently.

The 3D spatial structure and evolution of prominences or Coronal Mass Ejections (CMEs) could be investigated using observations collected simultaneously with two distant instruments (like Solar Terrestrial Relation Observatory A and B twin satellites), sufficiently separated to restore individual locations in 3D space by means of trigonometric parallaxes. Very valuable data concerning 3D evolution of the CME have already been collected with these satellites, but the amount of such data evaluated for prominences is unfortunately lower (\opencite{Panasenco2007}; \opencite{Slater2007}; \opencite{Bemporad2009}; \opencite{Liewer2009}).

Due to the unique optical characteristics of Multi-Channel Subtractive Double Pass (MSDP) imaging spectrographs (large field of view, possible short exposure times and high temporal cadences in scanning mode), collected data are convolutions of the 2D images and spectra (\opencite{Mein91a}, \citeyear{Mein91b}; \opencite{Rompolt93})). The method discussed in this work is suitable for analysis of the spatial motions and evolution of whole extended structures like big eruptive prominences visible over the solar limb or filaments on the solar disk. The method restores the true 3D trajectories of the prominence knots based on ground-based observations taken with a single telescope equipped with a MSDP imaging spectrograph. Our method allows to evaluate 3D trajectories of the prominence knots without any assumptions concerning the shape of trajectories or dynamics of the motion. The reconstructed trajectories of several blobs observed in three prominences are presented.

\section{Observational material and data processing}

The observational data were collected using the Large Coronagraph of the University of Wroc{\l}aw, equipped with the MSDP imaging spectrograph, installed in the Bia{\l}k{\'o}w Observatory. The coronagraph has a 51~cm entrance aperture, nearly 14.5~m effective focal length, although its real spatial resolution is limited by seeing only to about 1 arcsec. The MSDP spectrograph has a 9-channel prism-box creating 0.4~{\AA} steps in wavelength between consecutive "channels" forming spectro-images and defining in this way a spectral resolution of the collected data (\opencite{Rompolt93}). Consecutive channels cover exactly the same area on the Sun (325$\times$41 arcsec$^2$) but each of these is recorded in a slightly shifted waveband in relation to others.

Three series of prominence observations were selected from archive data covering the period 1996-2006 and data collected on August-September 2009.

The numerical reduction of the raw spectra-images was made using a standard MSDP software designed by Mein (\opencite{Mein91b}) and modified by Rudawy (\opencite{Rudawy1995}) and our own auxiliary codes. After numerical processing for each raw spectro-image we obtained a compound file containing 13 images in different wavelenghts in the neighborhood of the H$\alpha$ line ($\Delta \lambda = 0,$  \linebreak[3] $\pm 200 m$\AA,  \linebreak[3] $\pm 400 m$\AA,  \linebreak[3] $ \pm 600 m$\AA,  \linebreak[3] $ \pm 800 m$\AA,  \linebreak[3] $ \pm 1000 m$\AA, \linebreak[3] $\pm 1200 m$\AA) and a narrow field of view (defined by the entrance window of the spectragraph). Next, after numerical "sticking" of these images for each scan we got a compound file with 13 large-area images covering the investigated prominence. Usually we added dozen or so images to receive a single large-area image and we made several tens of scans for each observed prominence, reaching a temporal resolution of the order of 1 minute.

Due to the rotation of the field of view (the observations were made in coud\'e focus of the Large Coronagraph) and instabilities in the pointing of the telescope we de-rotated and translated numerically all images of the investigated prominence to the common reference frame of an arbitrary orientation and starting point. As reference structures we used sunspots, some stable structures on the limb (the place where they cross the solar limb) or stable feet of the prominences. Axes of the reference frames were arbitrarily oriented as follows: $x$ axis, from left to right, $y$ axis, from top to the bottom (both in the sky plane), $z$ axis, oriented from the observer to the Sun. The starting positions of the knots along the $z$ axis were not known, they were always assumed to be equal to 0.

\unitlength0.88pt

\begin{figure}[!t]
\begin{center}
   \fbox{\includegraphics[bb= 18 18 301 289, angle=0,
  width=0.95\textwidth,clip]{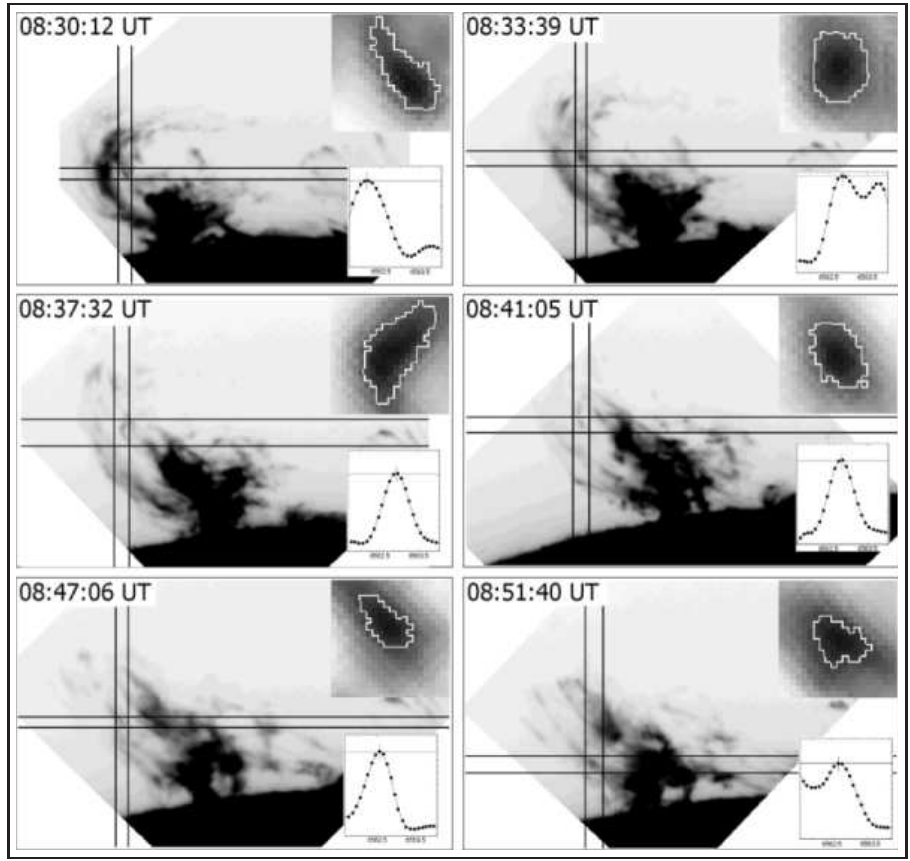}}
\caption{Images of an eruptive prominence observed on June 1, 2003 near active region NOAA 10375. The images were taken with the MSDP imaging
spectrograph in the center of H$\alpha$ line. The double-line cross surrounds the analyzed knot at its center. Zoomed images (top right),
mean H$\alpha$ profiles (bottom right corner) of the knot are presented in each panel.}
\label{zgestek03C}
\end{center}
\end{figure}

\unitlength1.0pt

Well defined knots of prominence material visible on at least five consecutive images were selected for further detailed analysis. An $x$-$y$ position of the blob was estimated on the consecutive images using the position of the centroid of the knot's emission (see Figure \ref{zgestek03C}) delimited by an arbitrary selected isophote (usually the isophote of 70 percent of the brightest knot's emission). The newest version of our software automatically points out knot positions on consecutive images, but results are examined by the observer to prevent bad tracking. In an automatic mode the isophotes in the range of 70\% - 90\% were automatically fitted to delimit the analyzed knot from other structures. The mean H$\alpha$ emission profiles of the knots were used to evaluate line of sight velocities of the material. As a result for each investigated knot we obtained a time series of its positions in the plane of the sky (expressed in kilometers) and its LOS velocities (in kilometers per second) at different times ($t_i$).

\begin{figure}[!h]
\begin{center}
  \fbox{\includegraphics[bb=  14 14 157 86, angle=0,
  width=0.95\textwidth,clip]{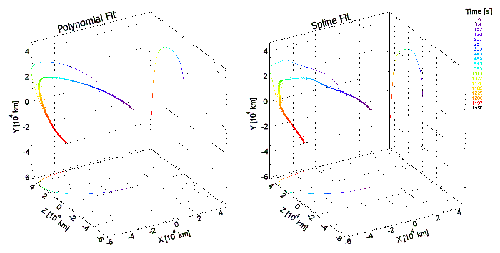}}
\caption{Three-dimensional trajectory of knot $A$ observed in the eruptive prominence on June 1, 2003. Left and right panels: polynomial approximation and spline interpolation of the observational data, respectively. Projections of the restored 3D track on the $X-Z$ and $Y-Z$ planes (both are perpendicular to the plane of the sky) as well as on the plane of the sky ($X-Y$) are plotted with thin lines. The time evolution is coded in color (violet - start, red - end of observations). Axes are scaled in kilometers.}
\label{polyspl}
\end{center}
\end{figure}

The temporal changes of the knot's spatial position and spatial velocity were calculated using an approximation of the obtained data $x(t_i)$, $y(t_i)$ and $V_{LOS}(t_i)$ with polynomials ($x_{poly}(t)$, $y_{poly}(t)$, $V_{LOS,poly}(t)$) and then by integration:
\begin{displaymath}
z_{poly}(t) = \int_{0}^{t}V_{LOS,poly}(t')\textrm{d}t'
\end{displaymath}
in order to obtain the translation along the $z$ axis.
Calculations were alternatively made by interpolation of the smoothed data ($x_{smooth}(t_i)$, $y_{smooth}(t_i)$, $V_{LOS,smooth}(t_i)$) with spline functions ($x_{spline}(t)$, $y_{spline}(t)$, $V_{LOS,spline}(t)$) and again by integration:
\begin{displaymath}
z_{spline}(t) = \int_{0}^{t}V_{LOS,spline}(t')\textrm{d}t'.
\end{displaymath}
The degrees of the polynomials used for the approximation were chosen for $x(t_i)$, $y(t_i)$ and $V_{LOS}(t_i)$ independently. In case of interpolation we smoothed raw data by the running mean to eliminate the influence of bad data. The time series of 3D coordinates $(x(t),y(t),z(t))_{poly}$ or  $(x(t),y(t),z(t))_{spline}$ directly define tracks of the knots in space. Figure \ref{polyspl} presents two versions of the trajectory of the same knot calculated using a polynomial approximation and a spline-function interpolation. Usually the trajectories calculated using a polynomial approximation were smoother than the trajectories calculated using a spline-function interpolation. While shapes of the majority of the magnetic loops observed in the solar corona can be well described by low-order polynomials, we decided to use polynomial-approximation for all trajectories presented in the next section of the paper.

\begin{figure}[!ht]
\includegraphics[bb= 0 320 623 1170,angle=90,width=0.95\textwidth,clip]{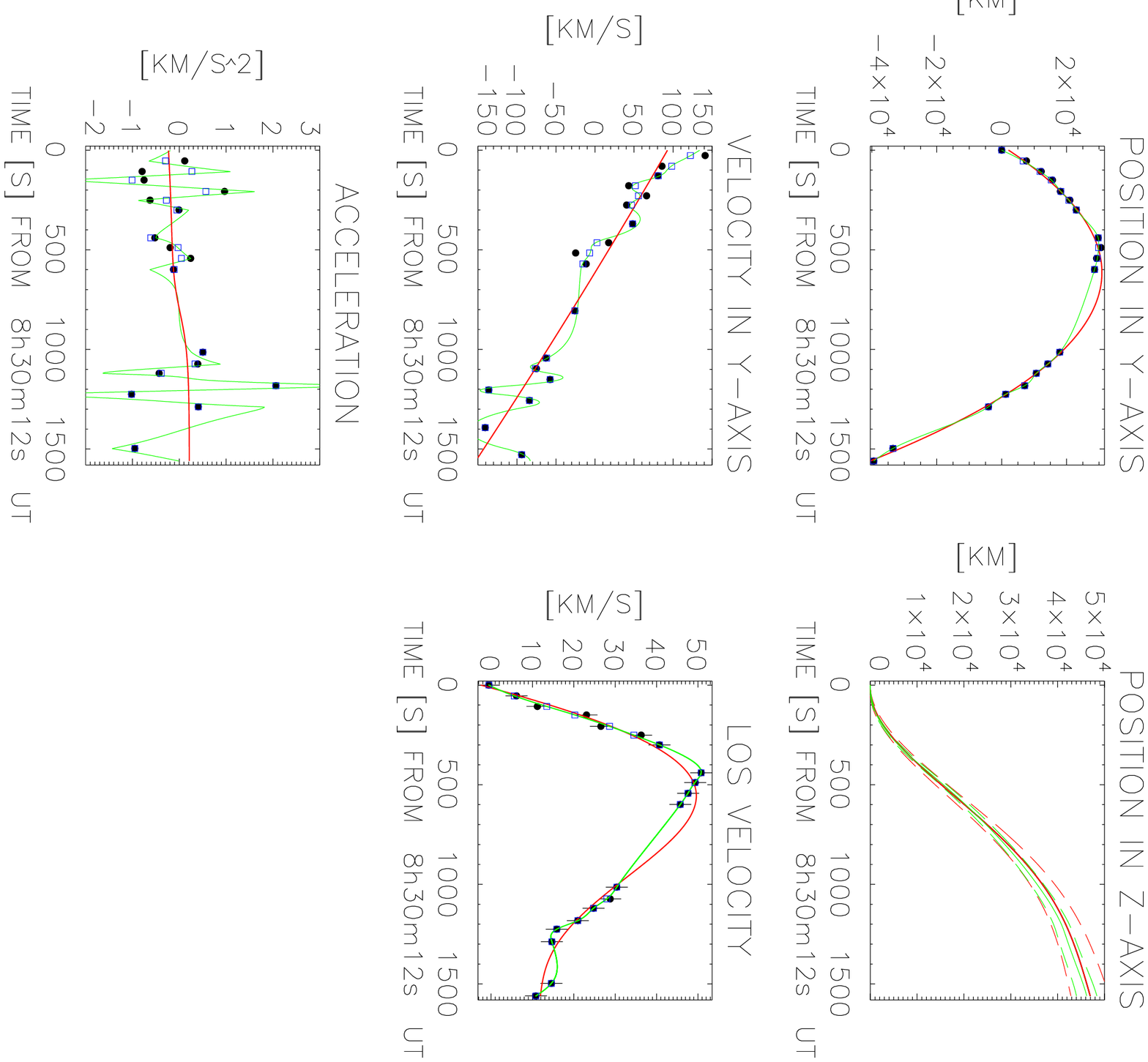}
\caption{Kinetic properties of the trajectory of the blob A observed in the eruptive prominence on June 1, 2003. Panels from top to bottom and from left to right: coordinates of the knot along $X$, $Y$ and $Z$ axes ($x$ and $y$ positions observed, $z$ position calculated), $v_x$, $v_y$ and $v_z$ velocities ($v_z$ observed, $v_x$ and $v_y$ calculated), full spatial velocity and acceleration along the trajectory. Raw data are marked with black points, smoothed data are marked with squares, solid red and green lines indicate spline and polynomial approximation, respectively. Error bars are marked by vertical bars.
In the top right panel the cumulative error of the integrated $z$ position is presented for the polynomial approximation (red dashed lines) and for spline interpolation (green dashed line).}
\label{kine}
\end{figure}

As was mentioned before, due to the obvious observational limitation we assumed arbitrarily for each knot $z(t=0)=0$, where $t=0$ was the beginning of the knot's observation period. In both methods we calculated also velocities along the $x$ and $y$ axes ($v_x$ and $v_y$ respectively) as first derivatives $x_{poly}(t)$ and $y_{poly}(t)$ (or $x_{spline}(t)$ and $y_{spline}(t)$) and full spatial velocities ($v=\sqrt{v_x^2+v_y^2+v_z^2}$) and accelerations along the path ($a=\frac{dv}{dt}$) (see Figure \ref{kine}). The obtained results are presented on Figures \ref{szpan05}, \ref{3d_nozero_2003} and \ref{3d_nozero_2009}.

Errors of the evaluated positions of the individual knots in the sky plane were estimated as the standard deviation of the positions of the reference structure in all analyzed images from which we obtained its mean value (the main source of the error was inaccuracies in co-alignment of the images). The estimated errors in $x$ and $y$ are less than 3~pixels (approximately 1000~km on the Sun). Errors of Doppler velocity measurements ($\Delta V$) were estimated for each knot separately using deviations of the individual measurements from their approximating polynomial. The error of the translation along the line of sight was assumed to be equal to the product of $\Delta V$ and the number of temporal steps used in the integration of the velocity curve from the first observed position of the knot. Thus, the error of the $z$ coordinate accumulated along the integration (see Figure \ref{kine}).

\section{Results}

\subsection{Post-flare loop system on July 14, 2005}

An extended post-flare loop system (PFLS) was observed from 9:22:54 to 9:46:03 UT on July 14, 2005, on the west solar limb over an active region NOAA 10786. The magnetic class of the region was $\beta \gamma$, it produced several tens of C and M class solar flares, according to \textit{GOES} (Geostationary Operational Environmental Satellite) classification. Five flares occurred shortly before the period of our observations: M9.1 \textit{GOES}-class flare at 05:57 UT and four C-class flares during the period lasting from 06:35~UT up to 08:54~UT. The observed PFLS appeared after the M-class flare. The Large Angle and Spectrometric Coronagraph (\textit{LASCO}) on board Solar and Heliospheric Observatory (\textit{SOHO}) satellite recorded a CME ejected from the closest vicinity of the active region at 05:40~UT.

\begin{figure}[!h]
\begin{center}
  \fbox{\includegraphics[bb= 14 14 302 196, angle=0,
  width=0.8\textwidth,clip]{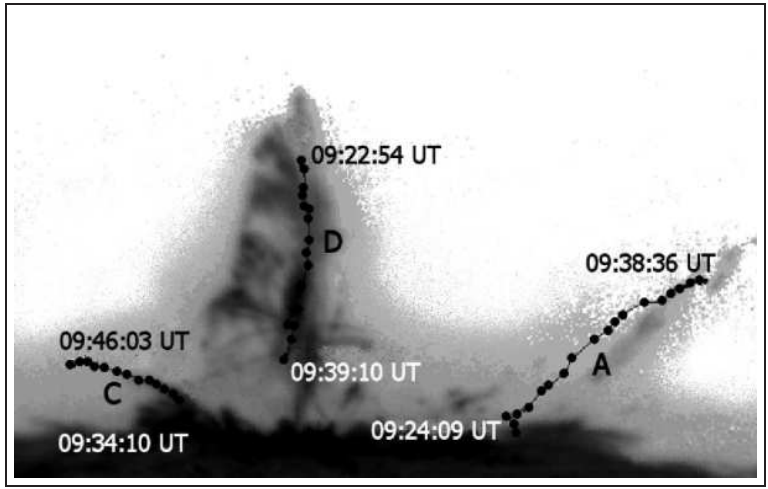}}
\caption{Observed positions of three knots (marked A, C and D) of the post-flare loop system observed on July 14, 2005. The image was taken with the MSDP imaging spectrograph in the center of the H$\alpha$ line at 09:36:02~UT.}
\label{xy05acd}
\end{center}
\end{figure}

\begin{figure}[!ht]
\begin{center}
  \fbox{\includegraphics[bb= 14 14 250 222, angle=0,
  width=0.95\textwidth,clip]{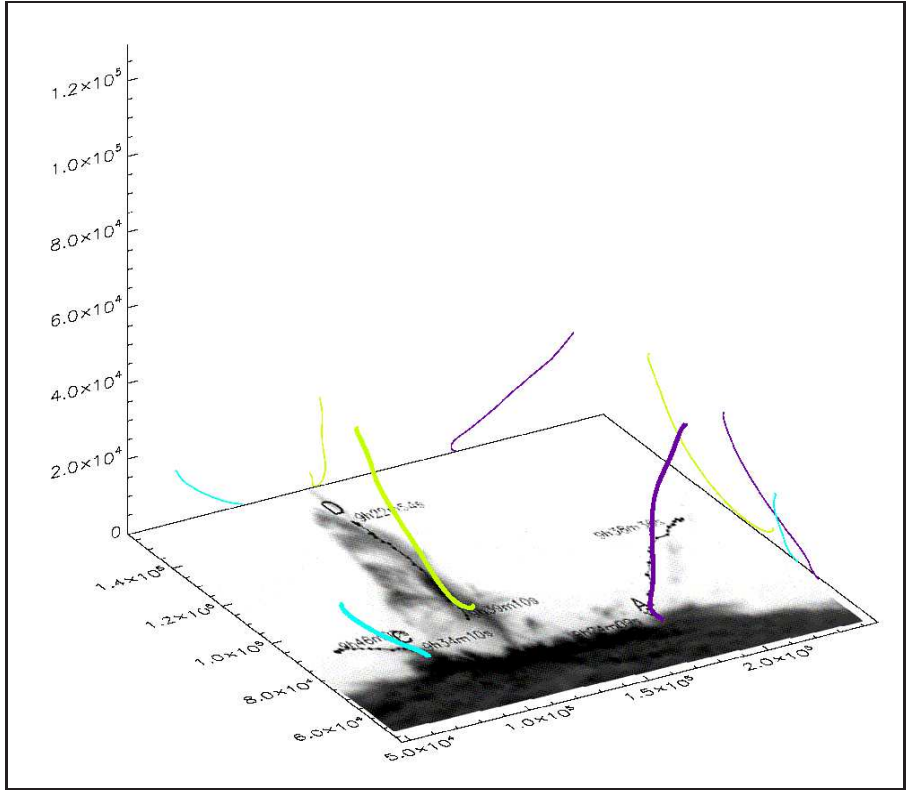}}
\caption{Restored 3D trajectories of the three knots (marked A, C and D) observed in the post-flare loop system on July 14, 2005. The tracks are presented in the 3D space and overlaid on the image taken with the MSDP spectrograph in the center of the H$\alpha$ line at 09:36:02~UT. Axes are scaled in kilometers. $Z$ axis (LOS) is vertical. Thick lines of various colors indicate 3D tracks of the individual knots (knot A - purple, C - light blue, D - light green, respectively). Projections of the restored 3D track on $X-Z$ and $Y-Z$ planes (both are perpendicular to the plane of the sky), as well as on the plane of the sky ($X-Y$) are plotted with thin lines. All trajectories are arbitrarily shifted along the $Z$-axis in such a way that the most distant position of each trajectory lies in the plane of the sky ($X-Y$).}
\label{szpan05}
\end{center}
\end{figure}

PFLS encompassed several loop-like structures and well separated knots, from which we selected three knots (marked A, C and D) for detailed analysis (see Table \ref{tab:23knots} for data details). Figure \ref{xy05acd} shows trajectories of the knots projected onto the sky plane, while Figure \ref{szpan05} presents restored 3D tracks of the same knots. The axes of the reference system are scaled in kilometers, LOS axis is vertical. The tracks are presented in the 3D space and overlaid onto an image of the PFLS taken in the $H\alpha$ line at 09:36:02~UT. Projections of the restored 3D tracks on the $X-Z$ and $Y-Z$ planes (both are perpendicular to the plane of the sky) as well as observed positions of the knots in the plane of the sky ($X-Y$) are also shown. For clarity $Z$-coordinates of starting positions of all knots are arbitrarily shifted in such a way that the most distant position on each trajectory lies in the plane of the sky ($X-Y$). 

The knot marked D was located inside the PFLS from 09:22:54 to 09:39:10 UT on 23 consecutive images. At the beginning of the observational period it had a positive Doppler velocity (~35 km s$^{-1}$), but afterwards it decelerated gradually to zero and next accelerated to ~40 km s$^{-1}$. Thus, its 3D track is bent out from the observer (see Figure \ref{szpan05}). Two other knots (marked A and C) were observed outside the PFLS (see Figures \ref{xy05acd} and  \ref{szpan05}).

\begin{center}

\begin{table}[!h]

\begin{tabular}{cccc} \hline
\centering
Knot & Start Obs. & End Obs. & No. of images \\ \hline
A &  9:24:09 &  9:46:03 & 33 \\
C &  9:34:10 &  9:46:03 & 20 \\
D &  9:22:54 &  9:39:10 & 23 \\ \hline

\end{tabular}
\caption{Knots observed in the post-flare loop system on July 14, 2005}
\label{tab:23knots}
\end{table}

\end{center}

\subsection{Eruptive prominence on June 1, 2003}

\begin{figure}[!h]
\begin{center}
  \fbox{\includegraphics[bb= 14 14 300 250, angle=0,
  width=0.95\textwidth,clip]{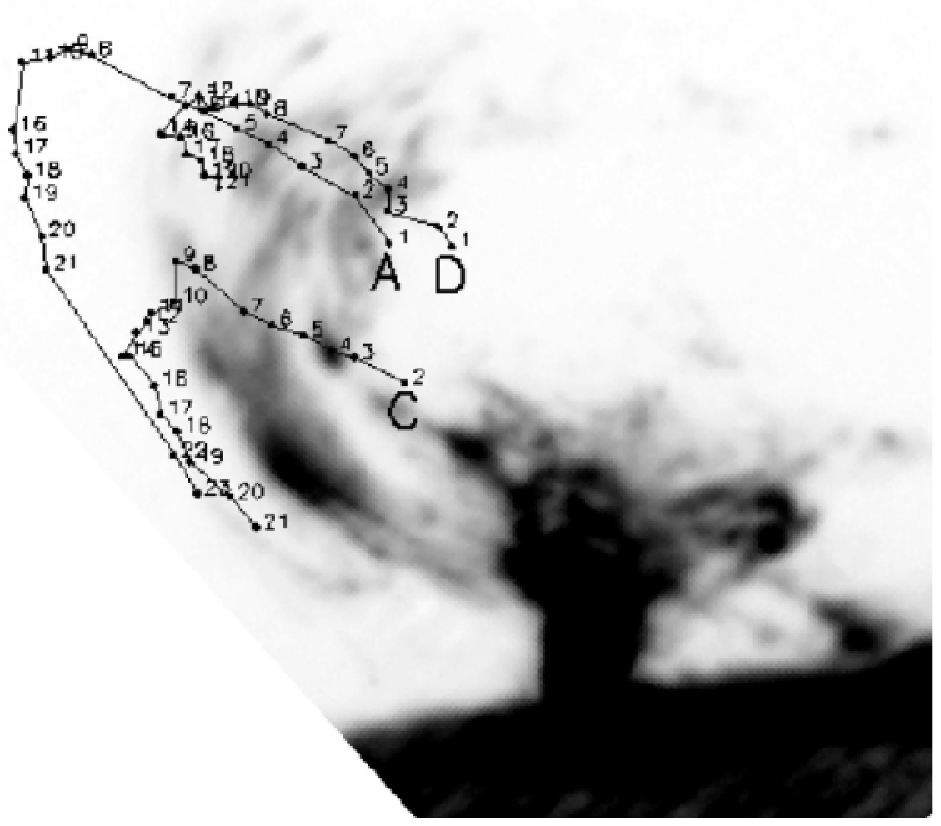}}
\caption{Observed positions of three knots (marked A, C and D) of the prominence observed on June 1, 2003. The image was taken with the MSDP imaging spectrograph in the center of the H$\alpha$ line at 8:32:42 UT.}
\label{3d_color}
\end{center}
\end{figure}

\begin{figure}[!h]
\begin{center}
  \fbox{\includegraphics[bb= 14 14 250 230, angle=0,
  width=0.95\textwidth,clip]{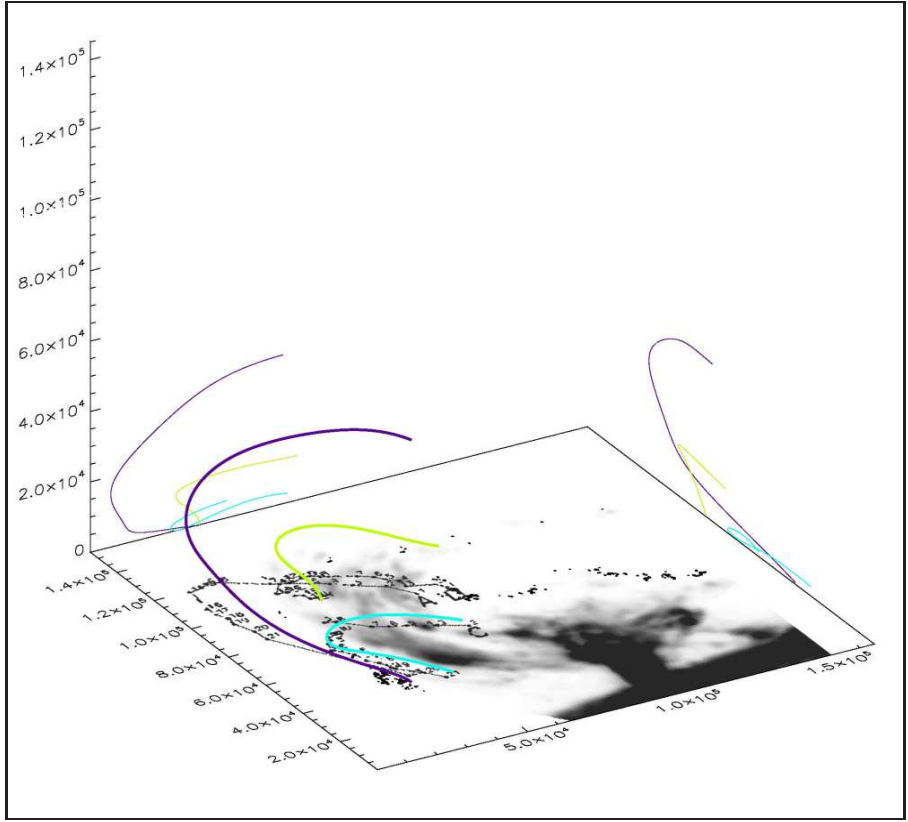}}
\caption{Restored 3D trajectories of the three knots (marked A, C and D) observed in the prominence on June 1, 2003. See Figure \ref{szpan05} caption for details.}
\label{3d_nozero_2003}
\end{center}
\end{figure}

The eruptive prominence was observed from 08:30:12~UT up to 09:13:57~UT near the active region NOAA 10375 on the east solar limb on June 1, 2003. An image of the prominence taken at 8:31:06~UT in the $H\alpha$ line is presented in Figure \ref{3d_color}, where trajectories of the selected knots projected on the sky plane are also shown. The restored 3D tracks of all knots are presented in Figure \ref{3d_nozero_2003}. Table \ref{tab:2003knots} summarizes the observations of the knots.

At the beginning of the observational period all knots had small radial velocities, but at the end of the observations knots marked A and C reached radial velocities of around 40~km/s. The recorded trajectories of the separated knots are combinations of motions of magnetic ropes as a whole and of their own motions within (along) the loops.

\begin{center}

\begin{table}[!h]
\begin{tabular}{cccc} \hline
Knot & Start Obs. & End Obs. & No. of images \\ \hline
A &  8:30:12 &  8:56:12 & 19 \\
C &  8:31:06 &  8:51:40 & 20 \\
D &  8:30:12 &  8:51:40 & 21 \\ \hline

\end{tabular}
\caption{Knots observed in the eruptive prominence on June 1, 2003}
\label{tab:2003knots}
\end{table}
\end{center}

\subsection{Quiescent prominence on September 23, 2009}

\begin{figure}[!h]
\begin{center}
  \fbox{\includegraphics[bb= 14 14 109 109, angle=0,
  width=0.6\textwidth,clip]{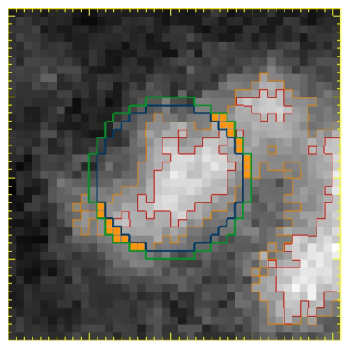}}
\caption{A portion of an image of a prominence, analyzed with an automatic code for tracking knots on consecutive images. Isophotes $70\%$ and $90\%$ are marked in orange and red respectively. Orange squares represent pixels located on the edge of the delimiting disk with signal larger than $L=70\%$ of maximum brightness of the knot on the evaluated image. In this example the red line inside the blue circle represents the knot's border.}
\label{blob_ex}
\end{center}
\end{figure}

A quiescent prominence was observed over the south-east part of the solar limb, close to active region NOAA 11026, on September 23, 2009. The prominence was observed by us from 08:07~UT up to 14:14~UT, we collected ~1000 scans over the whole prominence. The time cadence was between 15 and 18 seconds. Variable weather conditions caused several short gaps in observations.

We selected 6 well separated knots for further analysis (see Table \ref{tab:2009knots} for details). Due to a large amount of data we applied a code which tracks selected knots on consecutive images. On each image the code finds the minimum radius of the disk with center in the brightest pixel of the knot which has a percentage of pixels with signal larger than L and lower than E in its edge (between blue and green circles on Figure \ref{blob_ex}). In the second step the code finds the minimum (but larger than or equal to $L$) isophote level which lies inside the disk. This isophote is treated as the knot's border. The position of the knot is defined by the mean coordinates of all points inside the isophote weighted by the pixel signal values. $L$ value is chosen in the range $70-90\%$ of maximum brightness of the knot on the evaluated image. We found $E\approx 20\%$ to be suitable for analysis of the well separated knots, while $E\approx 50\%$ was better in more crowded fields. Despite the automatic procedure works quite well, each restored track also was examined by the observer.

\begin{figure}[!h]
\begin{center}
  \fbox{\includegraphics[bb= 14 14 242 212, angle=0,
  width=0.95\textwidth,clip]{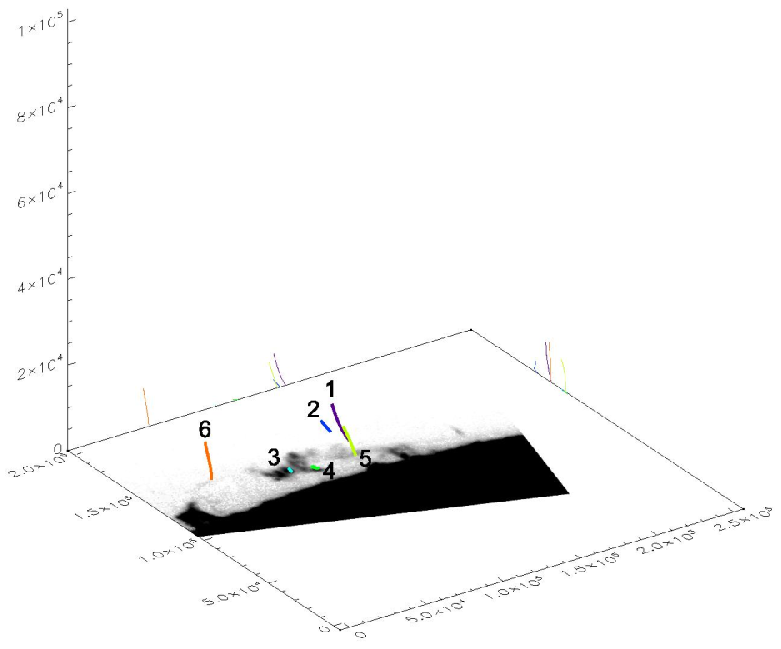}}
\caption{Restored 3D trajectories of the six knots (marked 1-6) observed in the prominence on 23 September, 2009. See Figure \ref{szpan05} caption for details.}
\label{3d_nozero_2009}
\end{center}
\end{figure}

The reconstructed tracks of the selected knots (marked from 1 to 6) are presented in Figure \ref{3d_nozero_2009}. Knots marked 1, 2, 5 and 6 had similar trajectories; they moved away from the observer, upward from the solar surface and from left to right in the sky plane. Knots 3 and 4 had small Doppler velocities and moved roughly in the sky plane.

\begin{center}

\begin{table}[!h]
\begin{tabular}{ccccc} \hline
Knot & Start Obs. & End Obs. & No. of images & Mean Doppler vel.\\ \hline
1 & 10:56:19 & 11:15:39 & 45 & 8.4 km s$^{-1}$\\
2 & 11:24:57 & 11:45:11 & 43 & 2.3  km s$^{-1}$ \\
3 & 13:58:26 & 14:00:54 & 9  & 4.1  km s$^{-1}$ \\
4 & 12:55:16 & 13:04:53 & 26 & 0.6  km s$^{-1}$ \\
5 & 11:22:07 & 11:30:45 & 23 & 15.6  km s$^{-1}$ \\
6 & 12:34:44 & 12:46:21 & 40 & 15.1  km s$^{-1}$ \\ \hline

\end{tabular}
\caption{Knots observed in the prominence on September 23, 2009}
\label{tab:2009knots}
\end{table}

\end{center}

\section{Discussion and conclusions}

In this paper we present a new method to restore the true spatial trajectories of prominence knots, based on ground-based observations taken with a single telescope equipped with a MSDP imaging spectrograph. The reconstructed trajectories of several knots observed in three prominences with the Large Coronagraph and MSDP spectrograph installed at Bia{\l}k{\'o}w Observatory are presented. Our method enables to evaluate 3D trajectories of the prominence knots without any assumptions about the shapes of the trajectories or the dynamics of the motion. Its only limitation is that starting positions of the knots along the line of sight could not be determined from the data and should be assumed to have any given value (most obviously to be equal to 0) or to be established from other data. Data like the one used in this work, combined with space-based observations, could be very useful to investigate the 3D evolution and dynamics of prominences, CME-prominence relations as well as overall evolution of huge magnetic systems.

%

\bibliographystyle{spr-mp-sola}

\end{article}

\end{document}